\documentclass{article}   

\usepackage{bm}
\usepackage{amsmath,amssymb}
\usepackage{graphicx}
\usepackage{listings}

\title{Keyboard Based Control of \\
Four Dimensional Rotations}

\author{Akira Kageyama\footnote{\texttt{kage@port.kobe-u.ac.jp} } \\[1em]
              Department of Computational Science,\\[0.5em]
               Kobe University, Kobe 657-8501, Japan
}
\date{}

\begin{document}

\maketitle

\abstract{
Aiming at applications to the scientific visualization of three dimensional simulations with time evolution,
a keyboard based control method to specify rotations in four dimensions is proposed.
It is known that four dimensional rotations are generally so-called double rotations,
and a double rotation is a combination of simultaneously applied two simple rotations. 
The proposed method can specify both the simple and double rotations by single key typings of the keyboard.
The method is tested in visualizations of a regular pentachoron in four dimensional space
by a hyperplane slicing.
}

\section{Introduction}

A three dimensional (3-D) computer simulation data with time evolution is
regarded as a data field defined in a four dimensional (4-D) space.
Spatio-temporal coherency in a 4-D field can be found when the 4-D data is 
appropriately rotated in the 4-D space 
to mix the spatial and temporal coordinates, before being mapped to 3-D or 2-D images.

The rotations in four dimensional  euclidian space are generally so-called double rotations.
A double rotation has two independent angles with two fixed planes that are absolutely perpendicular each other~\cite{cole1890rotations,Manning:1914}.
When one of the two rotation angles is zero, it is called simple rotation.
The study of 4-D rotations in the context of the visualization has a long history.
Mathematical basis of multi-dimensional rotations and its applications in graphics can be found in~\cite{hanson1995rotations}.
Since a 4-D rotation is represented by a composite of 3-D rotations,
sophisticated input methods for the 3-D rotations, 
such as ``rolling ball''~\cite{hanson1992rolling} or ``arcball''~\cite{shoemake1992arcball} can be used in 4-D rotations.
Various hardware have been used to implement the 4-D rotations,
from joysticks~\cite{banks1992interactive}, 
mice~\cite{hanson1999meshview},
haptic devices~\cite{Hanson:Multimodal2005}, 
flight-controller pads~\cite{sakai2011four},
head-tracked systems~\cite{sakai2007interactive},
and to modern touch screens~\cite{yan2012multitouching}.
Yan et al.~developed a multitouch interface of figure gestures for 4-D rotations~\cite{yan2012multitouching}.
Here we propose a much simpler approach to the 4-D rotations that is based on key typings of the keyboard,
rather than pursuing advanced input devices or methods.

Since a double rotation is equivalent to a product of two simple rotations~\cite{cole1890rotations},
it is reasonable to omit a specific user interface for the double rotations.
However, one has to invoke two commands for simple rotations simultaneously for a double rotation,
since an appearance of one step of a double rotation is not the same as that of two simple rotations if the two simple rotations are 
applied step by step.
Therefore, we assign in this paper a direct way to invoke the double rotations.
One of interesting features of the double rotations is that 
a continuously applied double rotation has no period of rotation when the ratio between the two angles is irrational.
It means that the rotated object never return to the original configuration even if the double rotation is continuously applied for ever.
A simple rotation in 4-D, on the other hand, has always a period $2\pi$ as in the rotations in 3-D.

The final goal of this study is to visualize 4-D data fields produced by time varying, 3-D numerical simulations.
As Woodring et al.~have shown~\cite{woodring2003high}, 
mapping a 4-D field into 3-D fields by hyperplanes in 4-D is effective to extract spatio-temporal coherency from the data.
A hyperplane in 4-D is written as
\begin{equation}\label{294729}
   c_0 + c_1 x + c_2 y + c_3 z + c_4 w = 0,
\end{equation}
where $c_i$ for $0\le i \le 4$ are constants.
In contrast to their ``hyperslice'' approach, in which general cases of the coefficients $c_i$ are considered,
we focus on the special cases of the type
\begin{equation}\label{295149}
   w = c_0.
\end{equation}
A slice of a 4-D data field $f(x,y,z,w)$ by the hyperplane of eq.~\eqref{295149} is just a 3-D field $f(x,y,z,c_0)$,
which is easily visualized by standard 3-D data visualization techniques.
Instead of considering general hyperplanes, 
we rotate the target 4-D data in the 4-D space under the fixed hyperplane.
From this reason, 
we investigate extrinsic 4-D rotations in which the target data is rotated about a fixed coordinates system.
The keyboard based control method proposed in this paper facilitates its user to keep track of the fixed planes.

We apply the proposed method to a visualization of a regular pentachoron which is the simplest figure in a 4-D space, but a 3-sphere.

Before we describe the proposed method in the next section,
here we briefly summarize some basic features of 4-D rotations.
In contrast to 3-D rotations, there is no rotation ``axis'' for a 4-D rotation.
Take, for example, 
a 3-D rotation around the $z$ axis, $R_z$, represented by the following matrix
that transforms a point $x_i = \left\{x_1,x_2,x_3\right\}=\left\{x,y,z\right\}$ to $x'_i$
\begin{equation}\label{292030}
R_z(\alpha) = \left(
\begin{matrix}
\cos\alpha &-\sin\alpha &0 \\
\sin\alpha &  \cos\alpha & 0\\
0 & 0 & 1 \\
\end{matrix}
\right),
\end{equation}
where $\alpha$ is the rotation angle.
It can be called a rotation in the $x$--$y$ plane.

Let us consider a 4-D rotation that transforms 
$x_i = \left\{x_1,x_2,x_3,x_4\right\}=\left\{x,y,z,w\right\}$ to $x'_i$:
\begin{equation}\label{103448}
R_{xy,zw}(\alpha,\beta)= \left(
\begin{matrix}
\cos\alpha &-\sin\alpha &0 & 0\\
\sin\alpha &  \cos\alpha & 0& 0\\
0 & 0 & \cos\beta & -\sin\beta \\
0 & 0 & \sin\beta & \cos\beta
\end{matrix}
\right),
\end{equation}
where $\alpha$ and $\beta$ are rotation angles.
In this 4-D case, the $x$ and $y$ coordinates are mixed with the angle $\alpha$,
and the $z$ and $w$ coordinates are mixed with the angle $\beta$.
A 4-D rotation does not have a fixed ``axis'',  but it has a two fixed ``planes'' that are perpendicular each other in 4-D space.
A double rotation is a commutable product of two rotations.
For example,  $R_{xy,zw}(\alpha,\beta) = R_{xy}(\alpha) R_{zw}(\beta)$, where
\begin{equation}\label{265228}
R_{xy}(\alpha) = \left(
\begin{matrix}
\cos\alpha &-\sin\alpha &0 & 0\\
\sin\alpha &  \cos\alpha & 0& 0\\
0 & 0 & 1 & 0 \\
0 & 0 & 0 & 1 \\
\end{matrix}
\right),
\end{equation}
and 
\begin{equation}\label{265232}
R_{zw}(\beta) = \left(
\begin{matrix}
1&0&0&0\\
0&1&0&0\\
0 & 0 & \cos\beta & -\sin\beta \\
0 & 0 & \sin\beta & \cos\beta
\end{matrix}
\right),
\end{equation}
that are called simple rotations.

\begin{figure}
  \centering
  \includegraphics[%
      height=0.6\textheight,%
       width=0.6\hsize,keepaspectratio]%
        {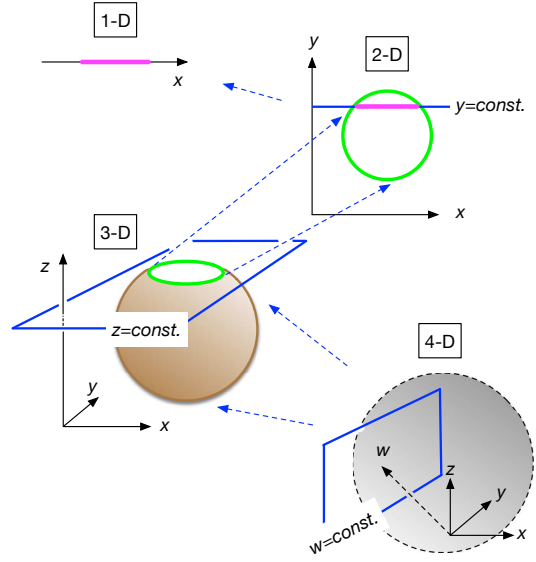}
  \caption{Slicing of an $N$-dimensional object makes $(N-1)$-dimensional object.}\label{290141}
\end{figure}

As mentioned above, a slice of a 4-D field $f(x,y,z,w)$ by $w=c_0=\text{const.}$  generates a 3-D field $f(x,y,z,c_0)$.
(Another major approach to the 4-D visualization is the projection of 4-D data into 3-D space.)
This slicing procedure is a natural extension from the lower dimensions as shown in Fig.~\ref{290141}:
The slicing of a 2-D object (green circle) in the $x$--$y$ space by a $y=\text{const.}$ line (blue) makes a 1-D object (magenta);
the slicing of a 3-D object (brown sphere) in the $x$--$y$--$z$ space by a $z=\text{const.}$ plane (blue) makes the 2-D object (green);
and the slicing of a 4-D object (gray) in the $x$--$y$--$z$--$w$ space by a $w=\text{const.}$ hyperplane (blue) makes the 3-D object (brown).

\section{Interactive 4-D Rotations by Keyboard}

\begin{table}[ht]
\begin{center}
\begin{tabular}{|c|c|c|}
\hline
Simple rotation 		& 	Axes numbers	& 	Key 		\\\hline\hline
$R_{x y}$	&	$1\times 2$	&	``2''	\\\hline
$R_{x z}$	&	$1\times 3$	&	``3'' 	\\\hline
$R_{x w}$	&	$1\times 4$	&	``4'' 	\\\hline
$R_{y z}$	&	$2\times 3$	&	``6'' 	\\\hline
$R_{y w}$	&	$2\times 4$	&	``8'' 	\\\hline
$R_{z w}$	&	$3\times 4$	&	``c'' 	\\\hline\hline
Double rotation		&	$x$'s counterpart  & 	Key		\\\hline
$R_{x y,z w}$ &   y			&	``y''	\\\hline
$R_{x z,y w}$ &   z			&	``z''	\\\hline
$R_{x w,y z}$ &   w			&	``w''	\\\hline
\end{tabular}
\end{center}
\caption{The key assignment for specifying the simple and double rotations.
The inverse rotation for each of them is realized by Shift keys.}
\label{254529}
\end{table}%

Here we propose a simple approach to 4-D rotations based on the keyboard and factorization.
We assign an integer to each of the four axes as follows:
\begin{equation}\label{255158}
x \rightarrow 1, 
\quad y \rightarrow 2,
\quad z \rightarrow 3,
\quad w \rightarrow 4.
\end{equation}
For a simple rotation $R_{x y}$, for example, 
two integers $1$ (for the $x$ axis) and $2$ (for the $y$ axis) are multiplied and the key ``2'' $[=1 \times 2]$ is pressed for the rotation.
Similarly, for rotations in $y$--$z$ plane, or $R_{y z}$ is invoked by 
the key ``6'' $[ = 2\: \text{(y-axis)} \times 3\: \text{(z-axis)}]$.
We use the hexadecimal notation; i.e., $R_{z w}$ is invoked by the key ``c'' $[=3\: \text{(z-axis)}\times 4\: \text{(w-axis)}]$.

While a key is pressed, the assigned rotation to the key is continuously applied to the 4-D object with a constant (usually small) angle $\theta_0$.
When the Shift-key is pressed, the inverse rotations, or negative angle rotations, are applied.
For example, the rotation $R_{w z}$, which is the inverse of $R_{z w}$,
is invoked by pressing Shift-c (or capital C) key.

The simple rotations are special cases when
one of two angles in the double rotations is zero.
There are three kinds of the double rotations under the rest frame:
$R_{x y, z w}(\alpha,\beta)$,  $R_{x z, y w}(\alpha,\beta)$, and $R_{x w, y z}(\alpha,\beta)$.
Focusing on the counterpart of the $x$ axis, 
we assign three keys ``y'', ``z'', and ``w'' for these double rotations, respectively.
The inverse rotations, such as $R_{y x, w z} = R^{-1}_{xy, zw}$,
are invoked again by pressing the Shift key.

The keys for the simple and the double rotations are summarized in Table~\ref{254529}.

The ratio between the two angles $\alpha$ and $\beta$ in the double rotations is also controlled
by the keyboard.
More precisely, when the ``k''  key is pressed,
the angle $\alpha$  is increased,
keeping $\alpha +\beta =\theta_0$.
Pressing the ``j'' key decreases the value of $\alpha$,
following the vi-editor's convention.
Either $\alpha$ or $\beta$ can be negative.

We will show, in the next section, examples of the 
proposed rotation method applied to a 4-D object visualized by 3-D slicing by a hyperplane.
The hyperplane is $w=c_0$ in a four dimensional $x$--$y$--$z$--$w$ space.
The $w$ coordinate of the slicing hyperplane is $c_0 = 0$ in the default settings.
The slice coordinate $c_0$ can be changed by pressing keys ``l'' or ``h''.
Following the vi-convention again;  the ``l'' key increases $c_0$ and ``h'' key deceases it.

The above keys are summarized in Table~\ref{281928}.

\begin{table}[ht]
\begin{center}
\begin{tabular}{|c|c|}
\hline
Key  		& 	Action			\\\hline\hline
``k''		&	Increase $\alpha$	\\\hline
``j''		&	Decrease $\alpha$	\\\hline
``l''		&	Increase $c_0$	\\\hline
``h''		&	Decrease $c_0$	\\\hline
\end{tabular}
\end{center}
\caption{The key assignments for parameters changes.
$\alpha$ is one of the two rotation angles for double rotations $R_{ij,kl}(\alpha,\beta)$. 
The incremental angle for $\beta$ is automatically determined by $\alpha+\beta=\theta_0$, where
$\theta_0$ is a constant angle for the simple rotations.
The parameter $c_0$ specifies the slicing hyperplane at $w=c_0$.
}
\label{281928}
\end{table}%

\section{Application to Visualizations of a Regular Pentachoron}

\begin{figure*}[t]
  \centering
  \includegraphics[%
      height=0.99\textheight,%
       width=0.99\hsize,keepaspectratio]%
        {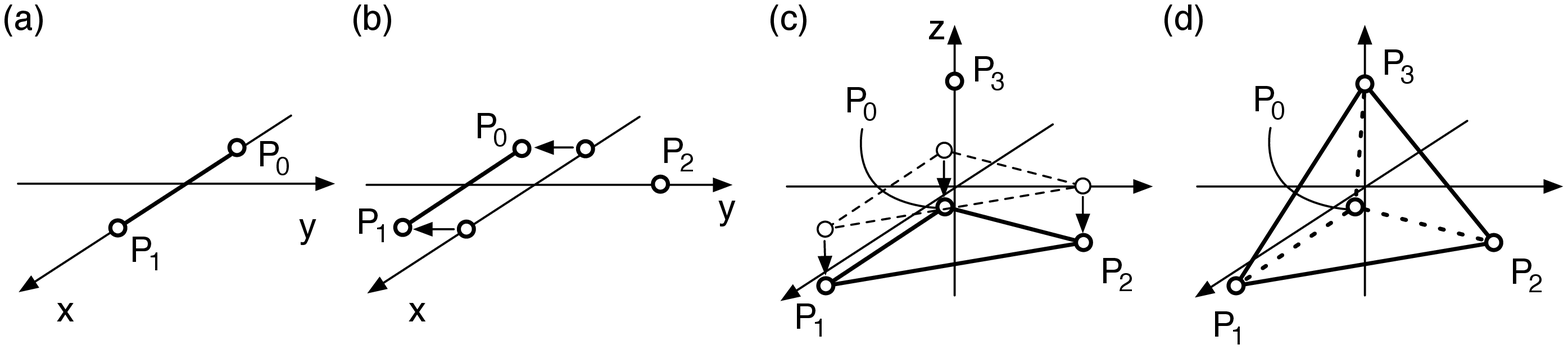}
  \caption{Construction of a four dimensional pentachoron with its center of gravity is on the origin.
  \label{081919}}
\end{figure*}

%
%

\subsection{A regular pentachoron}

As a sample application of the 4-D rotation,
we visualize a 4-D regular pentachoron.
In geometry, the analogs in $n$~dimensions of the polyhedra in 3-D space is called $n$-polytopes.
Especially in 4-D, a 4-polytope is called a polychoron.
There are six regular polychora, one of which is the regular pentachoron.
The regular pentachoron is represented as  $\{3,3,3\}$ in the Schl\"afli symbol,
indicating that it is a regular polychoron having three regular tetrahedra around each edge.
Since it has 5 congruent tetrahedra, a regular pentachoron is also called as a 5-cell regular polychoron.
The 4-D coordinates $(x,y,z,w)$ of the vertices of a regular pentachoron used in the next subsection are
\begin{align}
  \mathrm{P}_0 :& \left(-\frac{a}{2}, -\frac{a}{2\sqrt{3}},-\frac{a}{2\sqrt{6}},-\frac{a}{2\sqrt{10}}\right),     \label{082643a}  \\ 
  \mathrm{P}_1 :& \left( \frac{a}{2}, -\frac{a}{2\sqrt{3}},-\frac{a}{2\sqrt{6}},-\frac{a}{2\sqrt{10}}\right),     \label{082643b}  \\
  \mathrm{P}_2 :&  \left(     0,  \frac{a}{\sqrt{3}}, -\frac{a}{2\sqrt{6}},-\frac{a}{2\sqrt{10}}\right),     \label{082643c}  \\
  \mathrm{P}_3 :& \left(0,  0,  \frac{a\sqrt{3}}{2\sqrt{2}},-\frac{a}{2\sqrt{10}}\right),      \label{082643d}  \\
  \mathrm{P}_4 :& \left(0,  0,  0, \frac{2a}{\sqrt{10}}\right).    \label{082643e} 
\end{align}
The center of gravity of this regular pentachoron is on the origin.
These coordinates are constructed, for example, as follows.

Consider two points $\mathrm{P}_0$ and $\mathrm{P}_1$ on the $x$ axis whose coordinates are
$\bm{p}_0=-a/2$ and $\bm{p}_1=a/2$, respectively. 
See Fig.~\ref{081919}(a).
The distance between them is $a$,
and their center of gravity is on the origin.

We translate $\mathrm{P}_0$ and $\mathrm{P}_1$ in the negative $y$ direction so that
their coordinates are 
$\bm{p}'_0 =  (-a/2, -y_0/2)$, 
$\bm{p}'_1 = ( a/2, -y_0/2)$.
We place the third point $P_2$ on the positive $y$ axis at  
$\bm{p}'_2 = (0,y_0)$.
See Fig.~\ref{081919}(b).
The center of gravity of the triangle $\mathrm{P}_0\mathrm{P}_1\mathrm{P}_2$ is on the origin
because $\sum_{i=0}^2 \bm{p}'_i= 0$ for any $y_0$.
The distance between $\mathrm{P}_2$ and $\mathrm{P}_1$ is $a$ if $ y_0 = a / \sqrt{3}$.
From the symmetry,  $\left|\mathrm{P}_2-\mathrm{P}_0\right|=\left|\mathrm{P}_2-\mathrm{P}_1\right|$ .
The three points are on a circle of radius $r_2 = y_0$.
The figure $\mathrm{P}_0\mathrm{P}_1\mathrm{P}_2$ is a regular triangle with the edge length $a$.
A regular triangle is represented by $\{3\}$ with the Schl\"afli symbol.
It denotes a 3-sided regular polygon.

Keeping the above value $y_0 = a/\sqrt{3}$, 
we translate the regular triangle $\mathrm{P}_0\mathrm{P}_1\mathrm{P}_2$
in the negative $z$ direction so that their coordinates are
$\bm{p}''_0=  (-a/2, -y_0/2,-z_0/3)$, 
$\bm{p}''_1=  ( a/2, -y_0/2,-z_0/3)$, 
$\bm{p}''_2=  (     0,  y_0,  -z_0/3)$.
See Fig.~\ref{081919}(c).
We place the fourth point $P_3$ on the positive $z$ axis at  
$\bm{p}''_3 = (0,  0,  z_0)$.
The center of gravity of thus constructed tetrahedron $\mathrm{P}_0\mathrm{P}_1\mathrm{P}_2\mathrm{P}_3$
is on the origin 
because  $\sum_{i=0}^3 \bm{p}''_i= 0$ for any $z_0$.
The distance between $\mathrm{P}_3$ and $\mathrm{P}_i\: (0\le i \le 2)$ is $a$
if $z_0 = \sqrt{3/8}\, a$.
The four points $\mathrm{P}_i$ ($0\le i \le 3$) are located on a sphere:
with radius $r_3 = z_0$.
The figure $\mathrm{P}_0\mathrm{P}_1\mathrm{P}_2\mathrm{P}_3$ is, therefore, a regular tetrahedron with edge length $a$.
See Fig.~\ref{081919}(d).
A regular tetrahedron is represented as  $\{3,3\}$ with the Schl\"afli symbol,
indicating that it is a 
a regular polyhedron having three regular triangle faces around each vertex.

Finally, we construct a regular pentachoron in $x$--$y$--$z$--$w$ coordinate space, following the same procedure.
We translate the regular tetrahedron $\mathrm{P}_0\mathrm{P}_1\mathrm{P}_2\mathrm{P}_3$
in the negative $w$ direction so that their coordinates are
$\bm{p}'''_0=  (-a/2, -y_0/2,-z_0/3,-w_0/4)$, 
$\bm{p}'''_1=  ( a/2, -y_0/2,-z_0/3,-w_0/4)$, 
$\bm{p}'''_2=  (     0,  y_0,  -z_0/3,-w_0/4)$,
$\bm{p}'''_3 = (0,  0,  z_0,-w_0/4)$.
We place the fifth point $P_5$ on the positive $w$ axis at  
$\bm{p}'''_4 = (0,  0,  0, w_0)$.
The center of gravity of thus constructed pentachoron $\mathrm{P}_0\mathrm{P}_1\mathrm{P}_2\mathrm{P}_3\mathrm{P}_4$
is on the origin because  $\sum_{i=0}^4 \bm{p}'''_i= 0$ for any $w_0$.
The distance between $\mathrm{P}_4$ and other points $\mathrm{P}_i$  ($0 \le i \le 3$) is $a$
if $w_0 = \sqrt{2/5}\, a$.
The five points $\mathrm{P}_i$ ($ 0 \le i \le 4$) are located on a hypersphere (3-sphere) with
radius $r_4 = w_0$.
$\mathrm{P}_0\mathrm{P}_1\mathrm{P}_2\mathrm{P}_3\mathrm{P}_4$ is a regular pentachoron with edge length $a$.

The regular pentachoron with center of gravity in eqs.~\eqref{082643a}--\eqref{082643e} is thus constructed.

\subsection{A slice of a regular pentachoron by a hyperplane}

There are two major approaches to the visualizations of 4-D objects.
One is to apply orthographic or perspective projections in 4-D~\cite{noll1968computer,Sullivan:TheMathematicaJournal:1991,hanson1991visualizing,hoffmann1991some,Hanson:IeeeXplore:1993,hanson1999meshview,Chu:VCG:2009,sakai2011four}.
Another is to apply slicing with hyperplanes in 4-D.

In the slicing approaches, 
a 4-D object defined by a function $f(x,y,z,w)=0$ is sliced by a
hyperplane 
$c_x x + c_y y + c_z z + c_w w = c_0$, where $c_x$ etc.~are constants.
For example, the slice of hyperplane $w=0$ is a 3-D object $f(x,y,z,0)=0$
which can be observed in our 3-D $x$--$y$--$z$ space.

In contrast to the 4-D projection method,
the number of literatures on visualization of 4-D objects by the hyperplane slicing are relatively small.
Hausmann and Seidel developed a program to visualize four dimensional regular polytopes~\cite{Hausmann:ComputerGraphicsForum:1994}.
Their program can apply both the perspective projection and the hyperplane slices
of the six regular polytopes in 4-D.
In their program, however, the slice is very restrictive: 
The hyperplane is always along one of symmetry axes of the polytope.
A visualization of 4-D object by a hyperplane slicing was investigated
by Woodring et al.~\cite{woodring2003high} in which
general cases for arbitrary slicing hyperplanes are investigated.
Since the purpose of their study was to visualize time varying, three-dimensional scalar field,
the $w$ axis corresponds to time
and the 3-D slices were visualized by the 3-D volume rendering method.

Here we take the hyperplane slice approach to the 4-D visualization of the regular pentachoron.
The slicing hyperplanes are restricted to $w=0$ and its parallels $w=\text{const.}$
The 3-D slice is visualized by OpenGL in C++.

\begin{figure*}
  \centering
  \includegraphics[%
      height=0.7\textheight,%
       width=0.7\hsize,keepaspectratio]%
        {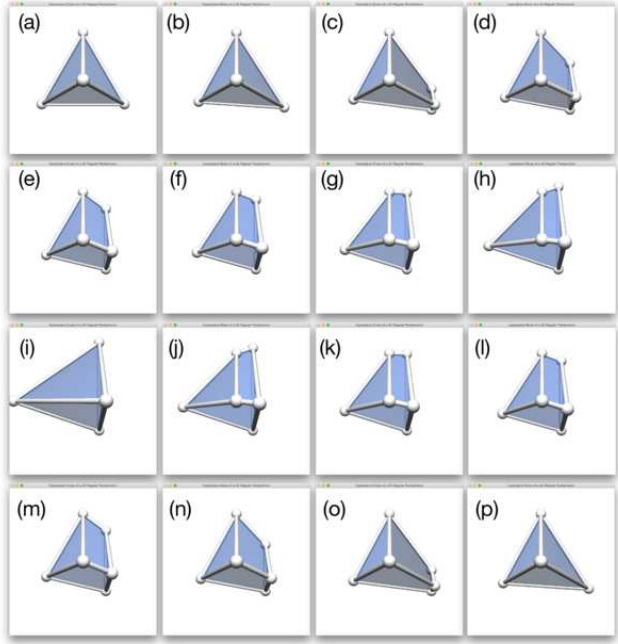}
  \caption{The 3-D slices of a 4-D regular pentachoron visualized by a slicing by hyperplane $w=0$.
  The regular pentachoron is continuously rotated by a simple rotation $R_{yw}(\theta)$ in 4-D
  by pressing ``4'' key of the keyboard.
  The rotation angle $\theta=\pi/16$.}\label{140157}
\end{figure*}

Figure~\ref{140157} shows a sample sequence of a simple rotation applied to 
the regular pentachoron $\mathrm{P}_0\mathrm{P}_1\mathrm{P}_2\mathrm{P}_3\mathrm{P}_4$ in eqs.~\eqref{082643a}--\eqref{082643e} with $a=2$.
The applied rotation is in the $x$--$w$ plane, or $R_{xw}(\theta)$, with $\theta=\pi/16$.
The panels from~(a) to~(p) show slices of the regular pentachoron by a hyperplane $w=0$.
It is a 3-D perspective view looking from $(x,y,z)=(0,0,5)$ to the negative $z$ direction.
The panel (a) is the initial view of the 3-D slice before rotation.
It is a regular tetrahedron in 3-D.
The vertices, edges, and faces are shown by balls, bars, and semitransparent polygons, respectively.
A vertex in 3-D is a slice of an edge in 4-D;
an edge in 3-D is a slice of a face in 4-D;
and a face in 3-D is a slice of a tetrahedron in 4-D.
The panels (b) to~(p) are slices by the same hyperplane $w=0$ of the pentachoron
when the simple rotation $R_{xw}$, by pressing ``4'' key, is continuously applied for every $\theta=\pi/16$ radian.
After applying the $R_{xw}(\theta)$ rotations for 32 times,
the regular pentachoron returns to the original configuration in 4-D, whose 3-D slice is shown in panel~(a).

\begin{figure*}
  \centering
  \includegraphics[%
      height=0.7\textheight,%
       width=0.7\hsize,keepaspectratio]%
        {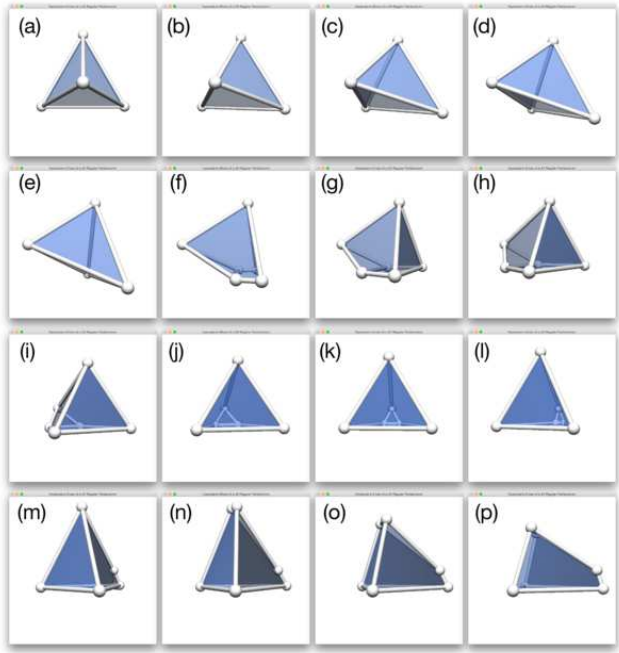}
  \caption{The 3-D slices of a 4-D regular pentachoron visualized by hyperplane $w=0$.
  The regular pentachoron is continuously rotated in 4-D by a double rotation $R_{xz,yw}(\alpha,\beta)$
  by pressing ``z'' key of the keyboard.
  The rotation angles in this figure are $\alpha=\pi\sqrt{2}/8\sqrt{3}$ and $\beta=1-\alpha$.
  The 4-D regular pentachoron never return to the initial configuration when the double rotations is continuously applied.}\label{033153}
\end{figure*}

Figure~\ref{033153} shows a sample sequence of a double rotation $R_{xz,yw}(\alpha,\beta)$,
where $\alpha=\pi\sqrt{2}/8\sqrt{3}$ and $\beta=1-\alpha$.
The panel (a) is the same slice as Fig.~\ref{140157}(a), before rotation.
The panels (b) to (p) are slices by $w=0$ when
the double rotation $R_{xz,yw}$, that is invoked by pressing ``z'' key, is continuously applied.
Since $\alpha/\beta$ is irrational, the 4-D regular pentachoron never return to the initial configuration, i.e., panel~(a),
in this double rotation.

\section{Summary}

Various input devices and methods have been investigated 
to specify the rotations in 4-D space.
We have proposed a simple approach to 4-D rotations based on the key typing of the keyboard.
Both the simple and double rotations can be specified by single key typings.
The rotations are supposed to be applied under a fixed frame of reference.
The extrinsic rotation helps the user to keep track of the orientation of the object in the 4-D space.
We can swiftly apply arbitral rotations to the 4-D object in a fully interactive way.
We can also smoothly change the slice coordinate $c_0$ through the key typing when the hyperplane slicing is applied for the 4-D visualization.

We have applied the proposed method to a visualization of a regular pentachoron
by the hyperplane $w=0$ and its parallels. 
The extension to the visualization of other regular polytopes in 4-D is straightforward.

One of the problems to be solved in the present approach to the 4-D visualization with
the combination of the interactive keyboard rotations and the hyperplane slicing
is that we cannot grasp a 4-D structure---at once---under the fixed hyperplane $w=c_0$.
A possible solution to this problem is to apply a lot of slices with slightly different $w=c_0$ coordinates at once and show them in a same window.
This is a kind of an extension of the computed tomography to 4-D.
We are developing a program based on this approach and the results will be reported in future.

\section*{Acknowledgment}
This work was supported by JSPS KAKENHI grant No.~20260052.



\begin{thebibliography}{10}

\bibitem{banks1992interactive}
{\sc Banks, D.}
\newblock Interactive manipulation and display of surfaces in four dimensions.
\newblock  { Proceedings of the 1992 symposium on Interactive 3D
  graphics} (1992), pp.~197--207.

\bibitem{Chu:VCG:2009}
{\sc Chu, A., Fu, C.-W.~W., Hanson, A.~J., and Heng, P.-A. .~A.}
\newblock GL4D: A GPU-based architecture for interactive 4D visualization.
\newblock { Visualization and Computer Graphics, IEEE Transactions} (2009), vol.15, no.6, 1587--1594.

\bibitem{cole1890rotations}
{\sc Cole, F.~N.}
\newblock On rotations in space of four dimensions.
\newblock { American Journal of Mathematics} (1890), 191--210.

\bibitem{hanson1992rolling}
{\sc Hanson, A.~J.}
\newblock The Rolling Ball.
\newblock { Graphics Gems III} (1992), 51--60.

\bibitem{hanson1995rotations}
{\sc Hanson, A.~J.}
\newblock Rotations for N-dimensional graphics.
\newblock { Graphics Gems V} (1995), 55--64.

\bibitem{Hanson:IeeeXplore:1993}
{\sc Hanson, A.~J., and Cross, R.~A.}
\newblock Interactive visualization methods for four dimensions.
\newblock { Proceedings of Visualization~'93} (1993), 196--203.

\bibitem{hanson1991visualizing}
{\sc Hanson, A.~J., and Heng, P.~A.}
\newblock Visualizing the fourth dimension using geometry and light.
\newblock {Proceedings of Visualization~'91} (1991),  321--328.

\bibitem{hanson1999meshview}
{\sc Hanson, A.~J., Ishkov, K.~I., and Ma, J.~H.}
\newblock Meshview: Visualizing the fourth dimension.
\newblock {Technical Report, Computer Science Department, Indiana University} (1999), 1--5.

\bibitem{Hanson:Multimodal2005}
{\sc Hanson, A.~J., and Zhang, H.}
\newblock { Multimodal exploration of the fourth dimension}.
\newblock Proceedings of Visualization~05 (2005), 263-270

\bibitem{Hausmann:ComputerGraphicsForum:1994}
{\sc Hausmann, B., and Seidel, H.-P.}
\newblock Visualization of regular polytopes in three and four dimensions.
\newblock { Computer Graphics Forum 13}, 3 (8 1994), 305--316.

\bibitem{hoffmann1991some}
{\sc Hoffmann, C.~M., and Zhou, J.}
\newblock Some techniques for visualizing surfaces in four-dimensional space.
\newblock { Computer-Aided Design 23}, 1 (1991), 83--91.

\bibitem{Manning:1914}
{\sc Manning, H.~P.}
\newblock { Geometry of four dimensions}.
\newblock Macmillan, New York, 1914.

\bibitem{noll1968computer}
{\sc Noll, A.~M.}
\newblock Computer animation and the fourth dimension.
\newblock  { Proceedings of the fall joint computer
  conference, part II\/} (1968), ACM, 1279--1283.

\bibitem{sakai2007interactive}
{\sc Sakai, Y., and Hashimoto, S.}
\newblock Interactive four-dimensional space visualization using
  five-dimensional homogeneous processing for intuitive understanding.
\newblock { Information and Media Technologies}, 2, (2007), 574--591.

\bibitem{sakai2011four}
{\sc Sakai, Y., and Hashimoto, S.}
\newblock Four-dimensional mathematical data visualization via embodied
  four-dimensional space display system.
\newblock { Forma} 26 (2011), 11--18.

\bibitem{shoemake1992arcball}
{\sc Shoemake, K.}
\newblock Arcball: A user interface for specifying three-dimensional
  orientation using a mouse.
\newblock { Proceedings of the Conference on Graphics Interface '92}
  (1992), vol.~92, pp.~151--156.

\bibitem{Sullivan:TheMathematicaJournal:1991}
{\sc Sullivan, J.~M.}
\newblock Generating and rendering four-dimensional polytopes.
\newblock {Mathematica Journal}, 1 (1991), 76--85.

\bibitem{woodring2003high}
{\sc Woodring, J., Wang, C., and Shen, H.~W.}
\newblock High dimensional direct rendering of time-varying volumetric data.
\newblock { Proceedings of Visualization~'03} (2003), 417--424.

\bibitem{yan2012multitouching}
{\sc Yan, X., Fu, C.~W., and Hanson, A.~J.}
\newblock Multitouching the fourth dimension.
\newblock { Computer}, 9 (2012), 80--88.

\end{thebibliography}

\end{document}